\documentstyle[11pt,newpasp,twoside,epsf]{article}
\def\edcomment#1{\iffalse\marginpar{\raggedright\sl#1\/}\else\relax\fi}
\reversemarginpar

\begin{document}
%%%%%%%%%%%%%%%%%%%
\title{Quasar Winds as Dust Factories at High Redshift}
\author{Martin Elvis, Massimo Marengo \& Margarita Karovska}
\affil{Harvard-Smithsonian Center for Astrophysics, \\
Cambridge Massachusetts, USA}

\begin{abstract}
Winds from AGN and quasars will form large amounts of dust, as the
cool gas in these winds passes through the (pressure, temperature)
region where dust is formed in AGB stars.  Conditions in the gas are
benign to dust at these radii.  As a result quasar winds may be a
major source of dust at high redshifts, obviating a difficulty with
current observations, and requiring far less dust to exist at early
epochs.
\end{abstract}

%%%%%%%%%%%%%%%%%%%%%%%%%%%%%%%%%%%%%%%%%%%%%%%%%%%%%%%%%%%%%%%%%%%%%%%%
\section{Introduction: Dust in High Redshift Quasars}

Dust is common in high redshift ($z$=4--6) quasars (Omont et al. 2001,
Priddey \& McMahon 2001), apparently implying that dust is widespread
by this epoch. This presents a puzzle: how can so much dust be formed
in the short time available? WMAP cosmology (Spergel et al. 2003) puts
the age of the universe at z=6 to be only 0.95~Gyr, and the age of
first reionization (and so the first stars) at an age of $\sim$0.2~Gyr
(z$\sim$20).  In our galaxy the primary source of dust lies in the
winds of red giant stars, specifically Asymptotic Giant Branch (AGB)
stars. Only stars with masses greater than 2~$M_{\odot}$ can reach the
AGB in the less than $\sim$1~Gyr available. Another source of dust is
thus required.

Normally this extra source is taken to be type~1 (massive star)
supernovae. Recent SCUBA results (Dunne et al. 2003) show that the
300~year old type~1 supernova remnant (SNR) Cas~A is rich in dust,
lending credence to this picture. However, the possible destruction of
SNR formed dust in their shocks may not be fully resovled. Here we
point out another, seemingly inevitable, path for dust formation up to
at least z=6: {\em quasar winds}.

The fate of outflowing broad emission line (BEL) quasar gas constantly
being ejected from quasars had previously not been considered. We
examined that fate (Elvis, Marengo \& Karovska 2002) in a manner that
applies to any model in which the BEL clouds move outward. We find
that dust creation is a natural consequence.

%%%%%%%%%%%%%%%%%%%%%%%%%%%%%%%%%%%%%%%%%%%%%%%%%%%%%%%%%%%%%%%%%%%%%%%%
\section{Quasar Winds}

Outflowing winds of highly ionized material are common in quasars [see
Arav, Shlosman \& Weymann (eds) 1997]. The gas emitting the prominent
Broad Emission Lines (BELs) may well participate in these winds.  
The BELs, which produce e.g.  Ly$\alpha$, H$\alpha$, H$\beta$ , CIV, NV
(e.g. Krolik 1999) are Doppler broadened to a few percent of the speed
of light ($\sim 3000-15,000$~km~s$^{-1}$). The gas producing these
lines lies at temperatures of 10$^4$~K and at the strikingly high
densities of 10$^9$-10$^{11}$~cm$^{-3}$ (Osterbrock 1989). These
densities are comparable to chromospheric values (Allen 1975, Korista
1999). The kinematics of this BEL gas - infalling, bound, or
outflowing - is not well established (Peterson 1997). Moreover, the
issue of how to prevent this high density gas from dispersing has been
problematic. Pressure confinement by a hotter surrounding medium would
seem straightforward and a 2-phase medium is even predicted for gas
irradiated by a hard quasar-like continuum (Krolik, McKee \& Tarter
1981). However this model appeared to suffer unsurmountable problems
(Mathews 1986). Most obviously the any BEL `clouds' would be ripped
apart by shear forces in less than one orbital time. As a result this
model was abandoned.

A BEL wind has advantages. Elvis (2000) showed how such a wind can
include the BEL gas as a cool phase in pressure equilibrium the warmer
(10$^6$~K) wind medium.  With this model a large number of other
puzzling features of quasar phenomenology seem to fall into place.
A wind in which both BEL gas and the hotter confining gas are
co-moving solves the cloud survival problem, and if the wind is
non-spherical also solves the problem of the confining medium being
Thomson thick, which cannot be the case as rapid X-ray variability is
essentially universal in AGNs. 

But a wind requires us to ask what happens to the gas as it flows
outward.  In any outflow model with the BEL region initially in
pressure equilibrium with a surrounding warmer medium, the divergence
of the outflowing warm wind (even if only at the sound speed of the
warm confining medium, $\sim$100~km~s$^{-1}$) will rapidly take the
system out of pressure balance. The BEL clouds will then begin to
expand, limited by their sound speed (initially $\sim$10~km~s$^{-1}$),
and will cool to below 1000~K, at which temperatures dust will form
{\em if} the pressure is still sufficiently high. Could quasar BEL gas
then be the source of quasar dust at high redshift?

%%%%%%%%%%%%%%%%%%%%%%%%%%%%%%%%%%%%%%%%%%%%%%%%%%%%%%%%%%%%%%%%%%%%%%%%%%%
\section{Smoking Quasars}

A full treatment of dust formation requires coupling the dust forming
medium hydrodynamics with the full set of dust condensation chemistry
equations (Sedlmayr 1997). Such an exercise is limited by our
knowledge of the highly non-linear dust condensation chemical paths,
and by the uncertainties related to the role of non-equilibrium
chemistry. We therefore use a simple comparison between AGN and cool
star atmospheres, to derive reasonable estimates for the conditions of
dust formation in the BEL clouds.

The general scenario for dust formation is based on the concept
of the ``dust formation window''. Effective dust condensation
seems to take place whenever a chemically enriched medium has a
sufficiently low temperature, and a large enough density, to
allow dust grain condensation. The amount of dust produced, the
chemical composition and the final size of the grains, depend on
the length of time over which the conditions remain favorable.

Figure~1 shows the dust formation window in pressure-temperature
space, for the chemical species that lead to the formation of dust in
an O-rich cool star circumstellar envelope. The thin solid lines mark
the phase transition region below which the most important dust
precursor molecules are formed.  The hatched area is the dust forming
region in the circumstellar envelope of a cool giant star. The region
is limited on the right by the thermodynamical path of a static
outflowing wind typical for an AGB star. The left side is obtained by
increasing the maximum pressure in the envelope by a factor 100, as
during the propagation of pulsational shocks in the atmospheres of
Long Period Variables. Dust formation in the envelopes of evolved
giants occurs in the region between the two tracks, below the phase
transition lines for each chemical species. A closely similar diagram
can be drawn for C-rich gas.

Figure~1 shows adiabats for BEL gas, assuming $\gamma = 5/3$. Clearly
they pass through the dust formation window for plausible initial
densities. The BEL gas enters the dust formation window after
expanding by a factor of three.  This process resembles that which
makes smoke in terrestrial settings. Quasars are then dusty because
they themselves create dust which, since carbon may be overabundant in
quasars, may resemble soot. Hence we called them `smoking quasars'.

%%%%%%%%%%%%%%%%%%
\begin{figure}
\begin{center}
\plotone{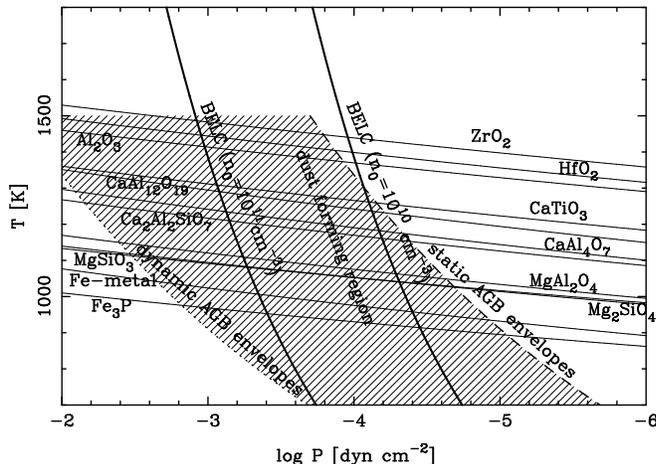}
\caption{Phase transition lines for O-rich dust precursor
molecules (adapted from Lodders \& Fegley 1999).  The hatched area is
the dust formation region in the circumstellar envelopes of evolved
cool giant stars, delimited by the two cases of static and dynamic
(pulsating) AGB atmospheres. The thick solid line is the path of BEL
clouds as they expand, for two different values of their initial
density.}
\end{center}
\end{figure}
%%%%%%%%%%%%%%%%%%

If quasar winds are driven by line radiation then the characteristic
instability of such winds suggests that shocks may be common. In stars
with `P~Cygni' profiles (OB stars and Wolf-Rayet stars) such winds
produce highly velocity-structured opacity (e.g. Stahl et al., 1993,
Owocki 2001). Quasars show similar structures (Turnshek 1988), and are
suggested by hydrodynamic modeling (Proga 2001). In AGB star winds
these intermittent pressure enhancements boost dust production
(figure~1), starting a nonlinear ``avalanche'' effect, as observed in
Miras and other Long Period Variables (H\"offner 1999). This suggests
that pressure fluctuations could greatly enhance the quasar dust
creation rate.

Can the dust survive the radiation field from the quasar BH?
Surprisingly a quasar wind provides a {\em more benign} environment
for dust than an AGB star wind, despite the quasar luminosity being
much higher than the $10^4$~L$_\odot$ luminosity of a typical giant
star. The large flux of energetic photons from the quasar continuum
might have overheated the newborn dust grains above their sublimation
temperature, delaying the occurence of dust condensation, perhaps
until it becomes impossible due to the ever decreasing gas density.
However, due to the much larger geometrical dilution in quasars, the
radiative flux reaching the BEL clouds interior is actually lower than
the stellar flux in the dust forming region of the giant's wind. For a
quasar of luminosity $10^{46}$~erg s$^{-1}$ the flux density 3~pc from
the quasar center is $\sim 10^7$~erg~cm$^{-2}$~s$^{-1}$ .  This is at
least one order of magnitude less than the $\sim 2 \cdot
10^8$~erg~cm$^{-2}$~s$^{-1}$ in the stellar case. So the dust
formation window of the BEL clouds is determined by the polytropic
expansion of the clouds gas, as we have assumed, and not by radiative
transfer, as in the case of circumstellar envelopes (Ivezi\'{c} \&
Elitzur 1997).

Other destruction mechanisms, such as dust sputtering by electrons and
ions, or chemical sputtering, are not very effective at the rather low
($T_K < 10^4$~K) kinetic temperature of the cloud medium.  Kinetic
sputtering by the surrounding medium becomes effective (Draine \&
Salpeter 1979) only for $T_K > 5 \times 10^5$~K. By the time the BEL
clouds start forming dust, they will be surrounded by a warm medium
having a temperature $T \sim T_0 u^{2(1-\gamma)} \sim 2 \times 10^5$~K,
which is already low enough to prevent their immediate destruction by
sputtering.

%%%%%%%%%%%%%%%%%%%%%%%%%%%%%%%%%%%%%%%%%%%%%%%%%%%%%%%%%%%%%%%%%%%%%%%%%%%
\section{Quasar Dust Masses}

The total amount of dust that is formed, however, depends on the time
spent by the clouds in the region favorable to grain condensation and
growth. This is given by the cloud expansion time-scale $\tau_{(BELC)}
= r_0/c_0 \sim 3$~yr, where $c_0\sim 10$~km s$^{-1}$ is the initial
sound speed of the cloud and $r_0 \sim 10^{14}$~cm (Peterson 1997)
their initial average size. This timescale is comparable with the time
spent by circumstellar grains in the region where dust growth is most
active, suggesting that the efficiency in dust production of BEL
region and late type giant winds may be similar.

The most luminous quasars, in which the highest dust masses are found
(Omont et al. 2001), have luminosities over 10$^{47}$erg~s$^{-1}$ and
so may have mass loss rates $>$10~M$_{\odot}$~yr$^{-1}$. Assuming the
same dust fraction as in AGB stars gives $\sim$10$^{7}$~M$_{\odot}$ of
dust over a nominal 10$^{8}$~yr lifetime. This approaches the amounts
detected but is still short by about an order-of-magnitude. However,
super-solar abundances are common in high redshift quasars, certainly
in carbon (Hamann \& Ferland 1999), providing a higher density of raw
material for the formation of precursor molecules. Dust condensation
is highly nonlinear (Frenklach \& Feigelson 1989), the amount of dust
formed in any particular quasar is hard to predict, but the dependence
on abundance is likely to be an $n^2$ process initially, allowing for
more dust to be produced than in our simple estimate.  As a result the
infrared emission of quasars may not require the normally assumed
large associated burst of star formation (Sanders et al. 1988).

%%%%%%%%%%%%%%%%%%%%%%%%%%%%%%%%%%%%%%%%%%%%%%%%%%%%%%%%%%%%%%%%%%%%%%%%%%%
\section{Consequences of Quasar Dust Creation}

Quasar winds provide an economical explanation for high z dust. 
Dust at z$\sim$4 is observed primarily from observations of quasars
and of sub-mm sources. If dust is everywhere and quasars illuminate
this pre-existing dust then the total dust mass at z=4 is large. 

If the only dust at high z is manufactured in quasars, then far less
dust is implied than if quasars simply illuminate pre-existing dust,
which would then need to be distributed among all high~z galaxies,
active or not.  The metals in the dust must, of course, have been
created by an earlier generation of massive stars, that go supernova
quickly. To establish high quasar BEL abundances, the star formation
must be local to the quasar nucleus, again suggesting a more limited
amount of high z star formation.

High~z counterparts to sub-mm SCUBA sources seem to imply enormous
starbursts, but conclusively distinguishing between a starburst and a
quasar as the root power source heating the dust seen by SCUBA is
extraordinarily difficult, leaving open the quasar option.

In this picture dust can be formed as soon as quasars form. Quasar
winds ($v>$1000~km~s$^{-1}$) readily exceed the escape velocity from a
galaxy (v$_{esc}\sim$200~km~s$^{-1}$), or even a rich cluster of
galaxies (v$_{esc}\sim$1000~km~s$^{-1}$).  Hence the dust they produce
will be ejected into the intergalactic medium.  Dust is an important
catalyst of star formation, as dust provides both shielding from
ambient ultraviolet radiation and an efficient cooling path for the
surrounding medium. The early creation of dust by quasars may then be
important for seeding star formation at slightly later times.

%%%%%%%%%%%%%%%%%%%%%%%%%%%%%%%%%%%%%%%%%%%%%%%%%%%%%%%%%%%%%%%%%%%%%%%%%%%
\section{Conclusions}

In summary, dust will inevitably be created from the free expansion of
quasar broad emission line clouds in an outflowing wind. The
association of dust with quasars is not then {\em necessarily} linked
with intense star formation around quasars, but may be a consequence
of the quasar activity itself. The creation of dust in quasar winds
may solve the puzzle of where the very first dust comes from, and in
doing so suggests an unexpected role for quasars in cosmology.

\acknowledgments{We thank Eric Feigelson for alerting us to the
problem of dust formation at high redshift, and both he and Fabrizio
Nicastro for valuable discussions. This work was supported in part by
NASA contract NAS8-39073 (Chandra X-ray Center).}

%%%%%%%%%%%%%%%%%%%%%%%%%%%%%%%%%%%%%%%%%%%%%%%%%%%%%%%%%%%%%%%%%%%%%%%%%%

%%%%%%%%%%%%%%%%

\begin{references}

%\reference Aguirre A., 1999, ApJ, 525, 583.

%\reference Aguirre A., Haiman Z., 2000, ApJ, 532, 28.

\reference  Arav N., Shlosman I., Weymann R.J. (eds), 1997, 
{\em Mass Ejection from AGN} (ASP, San Francisco), ASP Conf. 
Series, Vol. 128.

%\reference Cadwell B.J., Wang H., Feigelson E.D., Frenklach M.,
%1994 ApJ, 429,285.

%\reference  Clavel J., Wamsteker W. \& Glass I.S., 1989,ApJ, 337, 236.

\reference Draine B.T., Salpeter E.E., 1979, ApJ, 231, 77.

\reference Dunne L., Eales S., Ivison R., Morgan H., Edmunds M., 2003,
{\em Nature}, 424, 285.

%\reference Dwek E., 1986, ApJ 302, 363

%\reference Dwek E., Rephaeli Y., Mather J.C., 1990, ApJ. 350,104.

%\reference Edmunds M.G., Eales S.A., 1998, MNRAS 299, L29.

\reference Elvis M., 2000, ApJ, 545, 63 \& astro-ph/0008064.

\reference Elvis M., Marengo M. \& Karovska, 2002, ApJ, 567, L107 \&
astro-ph/0202002

\reference Frenklach M., Feigelson E.D., 1989, ApJ, 341, 372.

%\reference Frenklach M., Carmer C.S., Feigelson E.D., 1989 Nature, 339,196.

\reference Hamman F., Ferland G., 1999 ARA\&A, 487.

\reference H\"offner S., 1999, in proc. I.A.U. Symposium 191 on {\em
Asymptotic Giant Branch Stars}, T. Le Bertre, A. L\`ebre.

\reference Ivezi\'{c} Z., Elitzur M., 1997, MNRAS, 287, 799.

\reference Krolik J.H., 1999, {\em Active Galactic Nuclei}
(Princeton University Press, Princeton), p.365.

\reference Krolik J.H., McKee C.F., \& Tarter C.B., 1981, ApJ, 249, 422

%\reference Knapp G.R., 1985, ApJ, 293, 273

\reference  Lodders K., Fegley B., 1999, in proc. I.A.U. Symposium
no. 191 on {\em Asymptotic Giant Branch Stars}, T. Le Bertre,
A. L\`ebre and C. Waelkens eds., p. 279.

%\reference Marengo M., 2000, Ph.D. Thesis, International
%School for Advanced Physics, Trieste, Italy

\reference Mathews W.G., 1986 ApJ, 305, 187.

\reference Omont A., et al., 2001, A\&A, 374, 371.

\reference Osterbrock D.E., 1989, {\em Astrophysics of Gaseous
Nebulae and Active Galactic Nuclei} (Univ. Science Books, Mill
Valley).

\reference Owocki S.  2001, {\em Encyclopedia of Astronomy and
Astrophysics} (IoP, Nature, Bristol, London \& Oxford), vol.3
p. 2248.

\reference Peterson B.M., 1997, {\em An Introduction to Active
Galactic Nuclei} (Cambridge Univ. Press, Cambridge).

\reference Priddey R.S., McMahon R.G., 2001, MNRAS, 324, L17.

\reference Proga D., 2001, ApJ, 538, 684.

\reference Sabra B., Hamann F., 2001, ApJ, 563, 555.

\reference Sanders D.B., Soifer, B. T., Elias, J. H., Madore, B. F.,
Matthews, K., Neugebauer, G., Scoville, N. Z., 1988, ApJ, 325, 
74.

\reference Sedlmayr E., 1997, A\&SS 1997, 251, 103.

\reference Spergel D., et al., 2003, ApJS, 148, 175.

%\reference Sedlmayr E., 1994, in ``Molecules in the Stellar
%Environment'', U.G. J\o rgensen (ed.), Springer, Berlin, p. 163.

\reference Stahl O. et al., 1993, A\&AS, 99, 167.

%\reference Sunyaev R.A., Zel'dovich Y.B., 1972, Comments
%Ap. Space. Phys., 4, 173.

%\reference Turnshek D. 1988, {\em QSO Absorption Lines}
%eds. J.C.Blades, D.Turnshek, C.Norman [Cambridge:CUP], p.17.

%\reference  Whittet D.C.B., {\em Dust in the Galactic Environment} 
%(IoP, Bristol).


\end{references}
\end{document}